\documentclass[aps,prl,twocolumn,superscriptaddress]{revtex4}

\usepackage{graphicx}% Include figure files
\usepackage{dcolumn}% Align table columns on decimal point
\usepackage{bm}% bold math

\begin{document}

%\title{Polydisperion of functionalized nanoparticles affects observed multivalency effects}
\title{Origin of broad polydispersion in functionalized dendrimers and its effects on cancer cell binding affinity}

\author{Jack N. Waddell} 
\affiliation{Department of Mathematics, University of Michigan, Ann Arbor, Michigan 48109}
\author{Douglas G. Mullen}
\affiliation{Program in Macromolecular Science and Engineering, Ann Arbor, Michigan 48109}
\affiliation{Michigan Nanotechnology Institute for Medicine and Biological Sciences, Ann Arbor, Michigan 48109}
\author{Bradford G. Orr}
\affiliation{Michigan Nanotechnology Institute for Medicine and Biological Sciences, University of Michigan, Ann Arbor, Michigan 48109}
\affiliation{Program in Applied Physics, University of Michigan, Ann Arbor, Michigan 48109}
\affiliation{Department of Physics, University of Michigan, Ann Arbor, Michigan 48109}
\author{Mark M. Banaszak Holl}
\affiliation{Program in Macromolecular Science and Engineering, Ann Arbor, Michigan 48109}
\affiliation{Michigan Nanotechnology Institute for Medicine and Biological Sciences, Ann Arbor, Michigan 48109}
\affiliation{Program in Applied Physics, Ann Arbor, Michigan 48109}
\affiliation{Department of Chemistry, Ann Arbor, Michigan 48109}
\author{Leonard M. Sander}
\affiliation{Department of Physics, University of Michigan, Ann Arbor, Michigan 48109}
\affiliation{Michigan Center for Theoretical Physics,  University of Michigan, Ann Arbor, Michigan 48109}

\date{\today}

\begin{abstract}
Nanoparticles with multiple ligands have been proposed for use in nanomedicine. The multiple targeting ligands on each nanoparticle can bind to several locations on a cell surface facilitating both drug targeting and uptake. Experiments show that the distribution of conjugated ligands is unexpectedly broad, and the desorption rate appears to depends exponentially upon the mean number of attached ligands.  These two findings are explained with a model in which ligands conjugate to the nanoparticle with a positive cooperativity of $\approx 4kT$, and that nanoparticles bound to a surface by multiple bonds are permanently affixed.  This drives new analysis of the data, which confirms that there is only one time constant for desorption, that of a nanoparticle bound to the surface by a single bond.
\end{abstract}

\pacs{}
%\keywords{}

\maketitle
%\section{}
%\subsection{}

\begin{figure}
	\centering
		\includegraphics[width=3.5in]{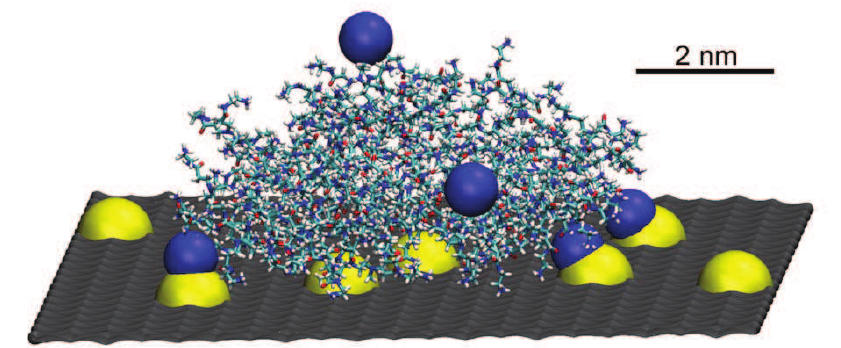}
	\caption{(Color Online) A dendrimer bound to a surface.  The dark spheres represent folic acid molecules  and the light spheres  folate binding protein, the folic acid receptor, on the surface.  Molecular dynamics simulation by C. Kelly.}
	\label{fig:Dendrimer}
\end{figure}

A dendrimer is  a branched polymeric nanoparticle with the topology of a Cayley tree \cite{Dykes-2001}; see Figure \ref{fig:Dendrimer}. We will be concerned here with dendrimers with a radius of about 5 nm  having $\approx 100$ termini that can be  functionalized by the conjugation of various endgroups and ligands.  These terminal groups  can be varied to tune solubility properties, and different ligands can be used to target and treat various cell pathologies.   \cite{Cheng2008, Jain2008} \cite{Venuganti2008}. 

Targeting ligands can be used to enhance the binding of the nanoparticle to specific receptors \cite{Stella2000}.  For example, epithieal cancer cells are known to overexpress folic acid receptors, so that folic acid attached to the dendrimer should target epithelial cancer, allowing chemotherapy agents also attached to the dendrimer to have high specificity \cite{Quintana2002}. For this application, it is important to understand the statistical distribution of the number of attached ligands and how this distribution affects binding to the cell surface. That is the subject of this paper. As we will see, this distribution is broad so that the fluctuation of the number of ligands is comparable to its mean. These large fluctuations are characteristic of physics at the nanoscale; this effect is often neglected. Further, the chemical reactions in question are always far from equilibrium. Analysis of the nanoparticle product from these reactions is in this case unique because detailed data is available on the ligand number distributions. This data allows us to explore effects, such as cooperativity in ligand conjugation during synthesis and multivalent enhancement of binding affinity, that would otherwise be unaccessible.

\begin{figure}
	\centering
		\includegraphics[width=3.5in]{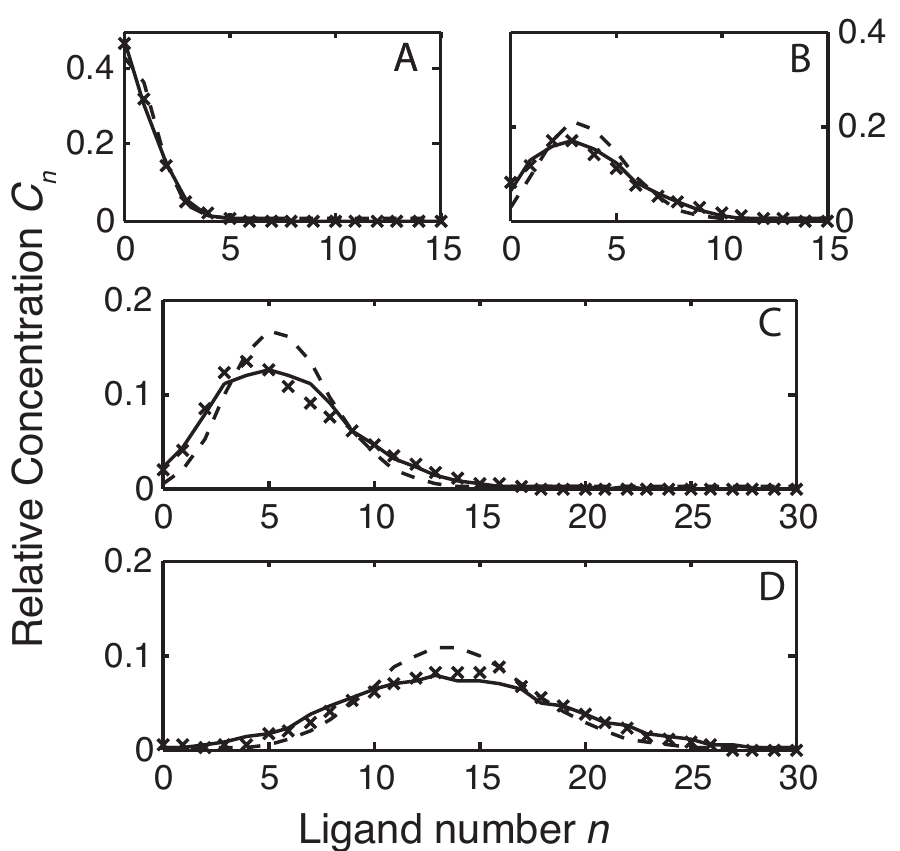}
	\caption{Distributions of dendrimers with $n$ bound ligands, for average ligand number $\bar{n}$ of (a) 0.9, (b) 3.7, (c) 5.8, and (d) 13.9.  The data points are HPLC data from ligation experiments that are chemically similar to the folic acid ligation.  The dashed line is the Poisson distribution and the solid line is the result of Monte-Carlo simulations with $E = - 3.7 \pm 0.1$kT.}
	\label{fig:dists}
\end{figure}

In our experiments measuring the distribution of ligand attachment \cite{Mullen2008} we conjugated varying amounts of  the ligand 3-(4-(prop-2-ynyloxy)phenyl)propanoic acid to the surface primary amines of a poly(amidoamine) dendrimer (G5 PAMAM; $(\mathrm{NH}_2)_{110})$. This ligand was chosen because its binding properties and steric constraints are similar to folic acid, and because it is amenable to separation by High Pressure Liquid Chromatography (HPLC).  Ligation takes place in a solution with $\bar n$ ligands available per dendrimer; the conjugation is by random attachment, and we assume that all the ligands attach. Let $C_n$ be the distribution of ligands on the dendrimers. $C_n$ is measured by HPLC.  We found that $C_n$ is very broad,  in fact \emph{broader than a Poisson distribution}; see Figure \ref{fig:dists}.  We attribute this effect to cooperativity, i.e., binding one ligand makes it easier to bind more ligands. 

The dendrimer  was experimentally determined to have approximately 110 free sites at the end of its (roughly) spherically arranged branches.  This is modeled as a 
11x11 triangular lattice with periodic boundary conditions.
We use a kinetic model of cooperative ligation with two parallel attachment paths.  In the free attachment path, a ligand attaches with a free energy barrier $E_0 = 0$.  In the cooperative path, a ligand attaches with free energy barrier $E$ to a site which has ligated neighboring sites.  
%When  $\bar{n}$ and the cooperativity are low enough a site can be assumed to have either 0 or 1 neighboring ligands.    In this case, rate equations can be written for $C_n$ \cite{Mullen2008}.  In the present case $\bar{n}$ and the cooperativity are too high to allow this approximation.

For large $\bar n$, two factors must be considered.  First, when a reaction site is proximate to a previously ligated site there is a possibility for a catalytic enhancement of the reaction rate.  The reaction occurs at the amide group of the ligand-nanoparticle bond, so the presence of multiple neighbors can increase the probability of a ligand 
attaching, say by orienting it properly \cite{Titskii-1970}. 
Second, multiple ligand neighbors crowd the site, sterically hindering ligation.  The data suggest that the hindrance is so strong that ligation does not occur when there is more than one ligand neighbor.  We write for the rate of attachment at a target with with $n_l$ ligated neighbors and $n_f$ free neighbors, where $n_l + n_f = 6$:
\begin{equation}
R = \omega_0 L(t)S(n_l) \big(n_l e^{-E/kT} + n_f \big) / \big(n_f + n_l \big). 
\label{eqn:rate}
\end{equation}
Here $\omega_0$ is a molecular time scale, $L(t)$ is the free ligand concentration, and $S(n_l)$ is a steric hindrance term that is equal to one if $n_l = \{0, 1\}$ and is zero otherwise.

We implemented a continuous-time, rejection-free Monte-Carlo simulation of the model with $N=10000$ dendrimers.  The reciprocal of $\omega_0$ is taken as the time unit. The only remaining variable is the free energy, $E$.
Simulations were run for values of average ligand number $\bar{n} = \{0.9, 3.7, 5.8, 13.9\}$, resulting in a best fit value of $E= -3.7 \pm 0.1$ kT.  The comparison between the monte-carlo simulations and the HPLC data are in Figure \ref{fig:dists}. This one-parameter fit to all the distributions is very satisfactory.

\begin{figure}
	\centering
		\includegraphics[width=3.5in]{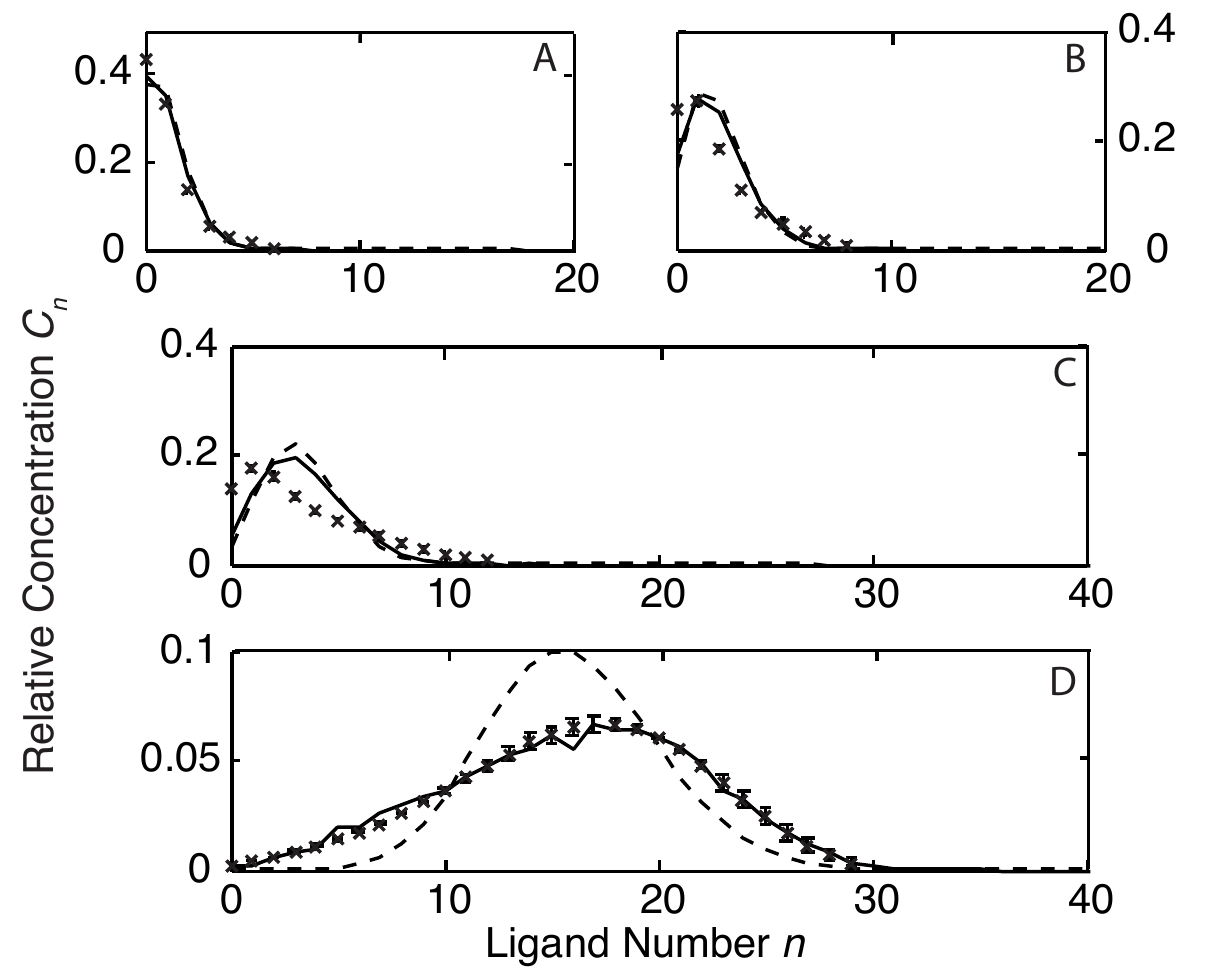}
	\caption{Distributions of dendrimers with ~ 68\% of sites acetylated and with $n$ bound ligands, for average ligand number $\bar{n}$ of (a) 1.0, (b) 1.9, (c) 3.4, and (d) 16.0.  The data points are HPLC data. 
The dashed line is the Poisson distribution and the solid line is the result of best-fit Monte-Carlo simulations with $E = - 5.0 \pm 0.2$kT.}
	\label{fig:distsAcetyl}
\end{figure}

We also used dendrimers with an average of 68\% of the active sites  blocked by the conjugation of acetamide groups (G5 PAMAM; $\mathrm{G5(Ac)}_{<78>}(\mathrm{NH}_2)_{<34>})$. We represented this by first using the model above to add acetamide groups using the acetamide-acetamide catalytic interaction free energy barrier $E_{aa}$ for $E$ in eqn \ref{eqn:rate} (though without steric hindrance; i.e., $S=1$). Then we added ligands, with steric hindrance occurring only between ligands due to the small size of acetamide. Now there is a new parameter, $E_a$, the ligand-acetamide interaction. Thus, the rate for ligand attachment at a site with $n_l$ ligand neighbors, $n_a$ acetamide neighbors, and $n_f$ free neighbors, where $n_l + n_a + n_f = 6$, is:
\begin{equation}
R = \omega_0 L(t) S(n_l) \frac{n_l e^{-E/kT} + n_a e^{-E_a/kT} + n_f}{n_f + n_l + n_a}, 
\label{eqn:rateExp}
\end{equation}
Because the catalyzing amide bond is present in both the acetamide- and the ligand-nanoparticle bond, we set $E = E_a = E_{aa}$. We have no new parameters, but still fit  the data quite well with one parameter; see Figure \ref{fig:distsAcetyl}. 

Having undersood the distribution of ligands on the nanoparticle we now turn to their adsorption and desorption from a protein-modified surface.  Surface Plasmon Resonance (SPR) can sensitively detect the amount of material adsorbed onto the surface.  In \cite{Hong2007}, SPR was used to determine the amount of folic acid (FA)-ligated dendrimers adsorbed on a surface covered with folate binding protein (FBP) as a function of time.  This gives the rate constants for adsorption, $k_a$, and desorption, $k_d$.  It was found that $k_d$ for the nanoparticles appeared to decrease rapidly with the mean number $\bar{n}$ of FAs. This was attributed to multivalent binding.

Multivalency, i.e. how multiple bonds between ligands on the nanoparticle and receptors on the surface effect binding is, in general, very complex.  However, if $C_n$ is broad,  apparent multivalent behavior may, in fact, be due simply to fluctuations in ligand number, as we will show. We explain the data of \cite{Hong2007} by supposing that nanoparticles are \emph{permanently} bound to the surface if they have two or more FA-FBP bonds.  Since nanoparticles with no FA are not bound at all, only those nanoparticles with precisely one FA-FBP bond contribute to $k_d$.  Thus for the time-scale of the SPR experiments the decrease in apparent $k_d$ with increasing $\bar{n}$ is due to the change in ligand distribution $C_n$ and the dilution of the dissociating material by the permanently bound material, not an enhancement in binding strength.    

\begin{figure}
	\centering
		\includegraphics[width=3.5in]{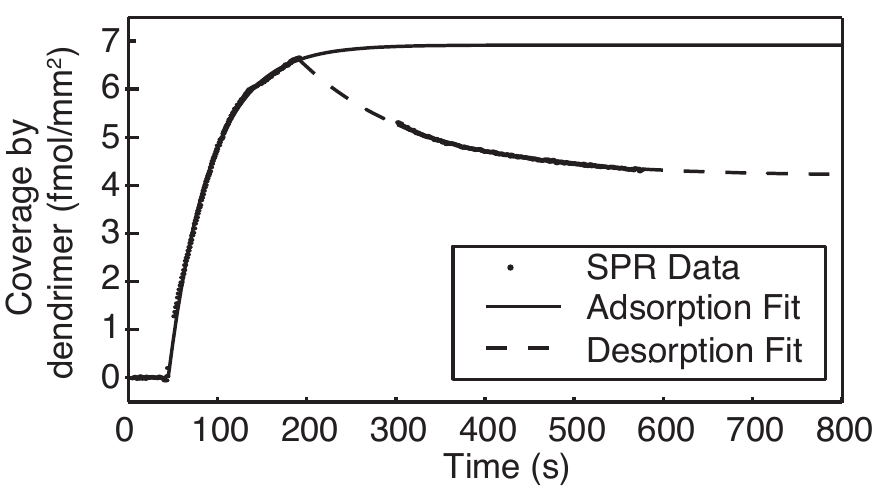}
	\caption{SPR measurement of nanoparticle coverage.  During adsorption (rising part of curve), buffer with  FA-bound dendrimer flows across the SPR chip.  During desorption, clean buffer flows; $k_d$ is obtained by fitting the desorption in the second phase, dashed line.}
	\label{fig:SPR}
\end{figure}

\begin{figure}
	\centering
		\includegraphics[width=3.5in]{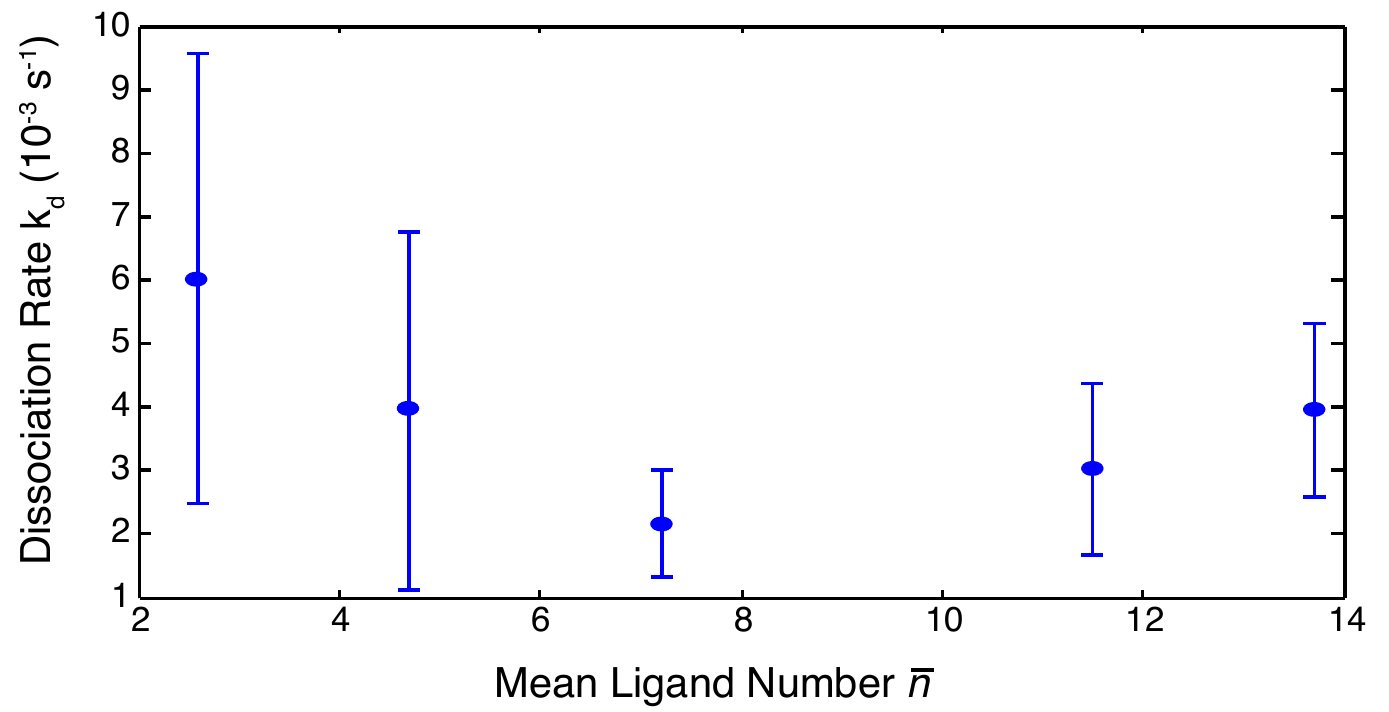}
	\caption{Dissociation rate $k_d$ versus $\bar{n}$, from exponential fits to the SPR data with a free horizontal asymptote.  The data are consistent with a rate independent of $\bar n$,  $k_d \approx 3 \cdot 10^{-3} \mbox{ s}^{-1}$, i.e.,  the rate of singly-bound nanoparticles dissociating from the surface.  }
	\label{fig:kds}
\end{figure}

This changes the analysis of the SPR data from that given in \cite{Hong2007}. In the standard multivalent model with nanoparticles with different  numbers of bonds to the surface, we expect the dissociation to involve many different rates.  However, if multiply bound dendrimers do not desorb over the course of the experiement, then there is one rate, that of a nanoparticle bound with a single FA-FBP bond. We reanalyzed the SPR data (see Figure \ref{fig:SPR}) to get new values of this $k_d$.  The results (see Figure \ref{fig:kds}) show a \emph{single} rate  $k_d \approx 3 \cdot 10^{-3} \mbox{ s}^{-1}$.  

\begin{figure}
	\centering
		\includegraphics[width=3in]{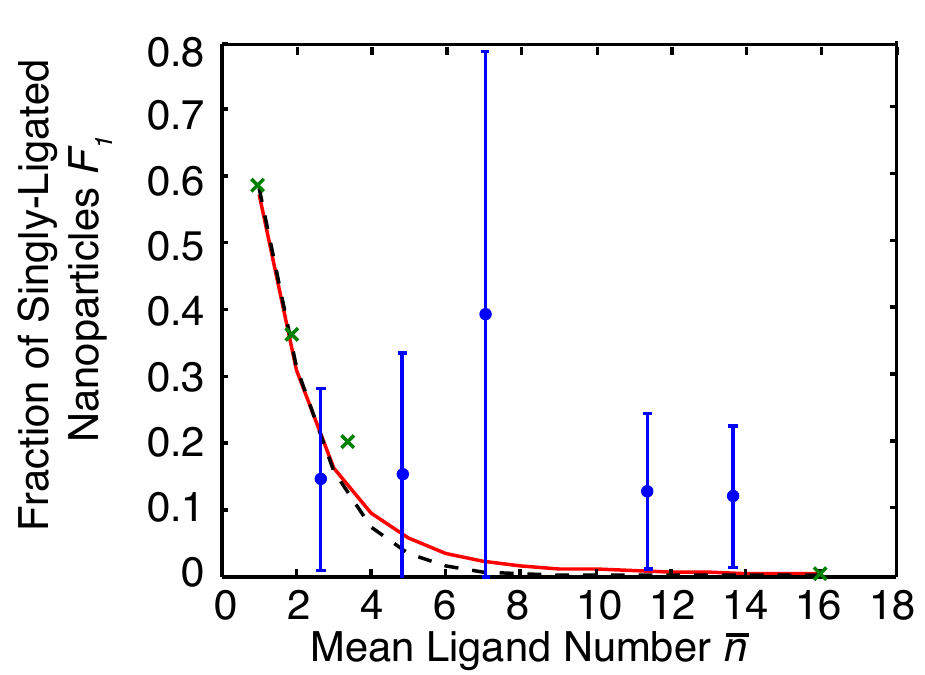}
	\caption{Fraction of singly-ligated nanoparticles vs.  $\bar{n}$. The  points are the fraction of residual material in the SPR experiment.  The  crosses are the fraction of singly-ligated nanoparticles as measured by HPLC.  The line is the fraction of singly-ligated nanoparticles $F_1$ simulated with the kinetic model with parameters $E = - 5.0$kT, which matched the best fit to the HPLC data.  The dashed line is $F_1$ for the Poisson case.}
	\label{fig:F1s}
\end{figure}

The number of singly bound nanoparticles is the fraction of bindable nanoparticles (i.e. with $n \ge 1$) that have \emph{exactly} one ligand.  Recall that if $C_n$ is Poisson then $C_1(\bar{n}) = \bar{n} e^{- \bar{n}}$.  However, since unligated nanoparticles are not bound at all, the fraction of bindable nanoparticles with one FA is $F_1(\bar{n}) = \bar{n} e^{- \bar{n}}/(1-e^{- \bar{n}})$.  This simple model is adequate at small average ligand number to explain the data (see Figure \ref{fig:F1s}).

As above, we use kinetic Monte-Carlo to find $C_n$ for FA on the dendrimers. 
From this, we estimate the fraction of singly-ligated nanoparticles, and thus what fraction of the material we expect to remain bound on the SPR surface.  The results are shown in figure \ref{fig:F1s}.  The SPR data is consistent with an approximately constant fraction of persistent material, whereas our simple model of singly-ligated dendrimers declines.  This is because the number of singly-ligated nanoparticles are an underestimate of the number of singly-bound nanoparticles.  A nanoparticle might have several ligands, but have only one accessible for binding due to steric effects.  Thus the theoretical limit is a lower bound for the dissociation, as is seen in figure \ref{fig:F1s}.  Even without these corrections, the simple theory fits the data if $\bar n$ is not too large. 

There are theories of multivalent interactions in monodisperse systems in the literature \cite{Huskens-2004, Diestler-2008}.  As we have pointed out, this sort of treatment is not necessary if time-scale separation exists between single-ligand interactions and multiple-ligand interactions.  This is the case when thermally released ligand-receptor bonds holding the nanoparticle to the surface are likely to reform before the remaining bonds break.  Based on the folic acid-folate binding protein binding energy estimates of Licata and Tkachenko \cite{Licata2008}, $\Delta E = 17 k_bT$, the time scale for the desorption of a doubly-bound nanoparticle is $1/k_d  \exp(\Delta E/k_b T) \approx 10^{10}\mbox{ s} \approx 300 \mbox{ years}$, a time inaccessible to the experiment and to biological processes.  Notably, the standard proprietary software used to analyze SPR data typically assumes that all adsorbed material will eventually desorb.  This assumption will result in an incorrect measurement in scenarios such as this in where there is a separation of time scales.

Monodisperse systems for which multivalent theories are posited may be rarer than anticipated; this is particularly the case for those systems in which dispersion is meant to be controlled by restricting the free ligand concentration (such as in \cite{Mullen2008, Ackerson-2006, Huo-2007}).  Many such multivalent systems are in fact composed of polydisperse particles with a broad ligand distribution.  Using a model such as ours, one may deconvolute data obtained from interesting polydisperse systems, even obtaining estimates for quantities such as the free energy of cooperativity for ligation.

Our goal in this paper is twofold: first we present an interesting mesoscopic system with its microscopic and macroscopic characteristics completely described by a simple statistical model. Also, we demonstrate that multivalent binding behavior observed in these and related chemical systems need not be explained by exotic interactions, but rather with simple physics combined with underlying distribution statistics. Our data gives distributions with HPLC and desorption rates from SPR, represent a rare case in which high quality information is available for both the small and the large scale of a mesoscopic system.  Our model describes both limits, estimates the free energy of cooperativity of about $-4$kT for the conjugation of ligands, and can predict ligand number distributions for other values of $\bar n$.  This also establishes the model as a tool for designing chemical syntheses to attempt to tune the dispersion, for example by correctly limiting the initial free ligand concentration or by quenching the reaction in progress at the appropriate time as the distribution evolves kinetically.

The model suggests that the strong steric hindrance prohibits newly attached ligands having more than one neighbor.  We have examined our simulation results, and we find that this leads to a much larger than chance occurrence of isolated pairs of ligands and, at larger $\bar{n}$, linear arrangements of ligands.  If true, this could have significant biological implications regarding that binding of a nanoparticle to a cell, since a ligand is likely to have a nearby neighbor.  The strong steric hindrance model also predicts that saturation of the nanoparticle of around ~50 ligands.  Hence the ligand distribution narrows for larger $\bar{n}$, at some point becoming \emph{narrower} than a Poisson distribution.

\begin{acknowledgments}
This project has been funded in part with Federal funds from the National Cancer Institute, National Institutes of Health, under Award 1 R01 CA119409, and with Federal funds from the National Science Foundation under Award DMS 0554587. 

\end{acknowledgments}

%\bibliography{theory}

\end{document}